%% file: 00_main.tex
\documentclass[fleqn,usenatbib]{rasti}

\usepackage{newtxtext,newtxmath}

\usepackage[T1]{fontenc}

\DeclareRobustCommand{\VAN}[3]{#2}
\let\VANthebibliography\thebibliography
\def\thebibliography{\DeclareRobustCommand{\VAN}[3]{##3}\VANthebibliography}

\usepackage{graphicx}	
\usepackage{amsmath}	
\usepackage{amsfonts}
\usepackage{bm}
\usepackage{subcaption}

\usepackage{xcolor} 
\usepackage{soul}

\graphicspath{{./images/}}
\DeclareGraphicsExtensions{.png,.pdf}

\title[All-Sky Imaging with Vector Sensors]{All-Sky Imaging with Vector Sensor Interferometry}

\author[Kononov and Knapp]{
Ekaterina Kononov,$^{1}$\thanks{E-mail: ekaterina.kononov@ll.mit.edu}
and Mary Knapp$^{2}$
\\
$^{1}$Lincoln Laboratory, Massachusetts Institute of Technology, Lexington, Massachusetts, USA\\
$^{2}$Haystack Observatory, Massachusetts Institute of Technology, Westford, Massachusetts, USA
}

\date{Accepted August 8, 2025. Received February 20, 2025; in original form ZZZ}

\pubyear{\the\year{2025}}

\begin{document}
\label{firstpage}
\pagerange{\pageref{firstpage}--\pageref{lastpage}}
\maketitle

\begin{abstract}
Radio astronomy observations at frequencies below 10~MHz could provide valuable science, such as measuring the cosmic dark age signal in the redshifted 21~cm hydrogen absorption line, detecting exoplanetary auroral emissions which lead to inferences about magnetic fields and atmospheres, and characterizing the effects of solar wind and coronal mass ejections on the magnetospheres of solar system planets. Despite their value, few resolved measurements in the sub-10~MHz band have been made. At frequencies below 10~MHz, the Earth's ionosphere reflects, attenuates, and distorts radio waves, making radio astronomy in this band possible only from space. However, a space-borne array would need thousands of electrically-small antennas to reach the sensitivity required for detecting faint astronomical signals, and it would need to be positioned far from the Earth to reduce the impact of Earth-based radio interference. Using more efficient antennas would minimize the number needed, and using antennas that are more resilient to interference would reduce the required distance from Earth. To this end, we consider constructing a low frequency array out of vector sensor antennas. These advanced antennas consist of three orthogonal dipole and three orthogonal loop antennas with a common phase centre. Their benefits include direction-finding and polarimetric capabilities, but they have not been considered for this application previously. We show that vector sensors can provide four times more Fisher information during interferometry than tripoles, simpler antennas that are commonly considered for space applications. We also present an all-sky imaging simulation to demonstrate a spherical harmonic imaging technique.
\end{abstract}



\input{01_intro}
\input{02_equation}
\input{03_spherical}
\input{04_coverage}
\input{05_sim}
\input{06_conclusion}

\section*{Acknowledgements}
The authors acknowledge the MIT SuperCloud and Lincoln Laboratory Supercomputing Center for providing computing resources that have contributed to the results reported in this paper. The authors declare no conflicts of interest. This work was funded by NASA grants 80NSSC23K0585 and 80NSSC24K1236. \copyright~2025 Massachusetts Institute of Technology.

\section*{Data Availability}

This work used publicly available data, the 2014 reprocessed Haslam sky map, which is available from NASA at
\url{https://lambda.gsfc.nasa.gov/product/foreground/fg_2014_haslam_408_get.html}.


\bibliographystyle{rasti}
\bibliography{references}





\bsp	
\label{lastpage}
\end{document}

%% file: 01_intro.tex
\section{Introduction}
The low-frequency sky below about 10~MHz (30~m) is obscured by the Earth's ionosphere, the layer of charged particles above the neutral atmosphere. Single spacecraft have made measurements in this band, e.g., \textit{RAE-2} \cite[]{Alexander1969} or \textit{Cassini} \cite[]{Gurnett2004}, but could not achieve high or even moderate angular resolution because a telescope's resolution $\theta$ is set by $\theta = \lambda/D$, where $\lambda$ is the wavelength and $D$ is the telescope diameter. For wavelengths that range from tens of meters to kilometres, a telescope must be hundreds of meters to many kilometres in diameter for even moderate resolution. Interferometry, a technique that combines signals from many spatially separated receivers to form a large virtual telescope, is ideally suited to long-wavelength astronomy.

Several compelling science questions could be answered with a long wavelength observatory. For instance, observations of auroral radio emissions and solar radio bursts can help scientists improve space weather forecasting and develop strategies to mitigate its effects on satellite operations, power grids, and communication networks \cite[]{Erickson2018,Kasper2021}. Second, cosmologists use observations of the redshifted 21~cm hydrogen absorption line to study the early universe. Deeply redshifted signals ($z=30-1000$) contain clues about the 'cosmic dark ages', an era shortly after the Big Bang but before the first stars and galaxies formed \cite[]{Jester2009,koopmans_peering_2021}. An orbital observatory would enable observations of those signals which fall near or below 10~MHz. And third, cyclotron maser radio emissions from exoplanetary aurorae encode information about magnetic fields \cite[]{kao2023resolved} and lead to inferences about their compositions and atmospheres \cite[]{zarka_magnetospheric_2015}. The emissions of many terrestrial planets, if present, are predicted to be below Earth's ionospheric cut-off \cite[references therein]{Lynch2018}, so a space-borne observatory would significantly improve observability of these signals.

Several low frequency interferometer space missions are currently being built. \textit{SunRISE} consists of six CubeSats in a nearly geosynchronous orbit, which will observe solar radio bursts \cite[]{Kasper2021}. \textit{AERO-VISTA} is a pair of CubeSats that will observe Earth's auroral emissions with vector sensor antennas \cite[]{Erickson2018, lind_aero_2019}. \textit{CURIE} is a pair of CubeSats in low Earth orbit tasked with studying solar radio bursts \cite[]{Sundkvist2016}. If successful, these missions will lay a foundation for future space-based radio interferometers. 

There has been steady interest in a low frequency space-borne array for several decades, and roughly every decade a new feasibility study is performed, considering technologies which matured since the previous study. One such recent study, called The Great Observatory for Long Wavelengths (GO-LoW) \cite[]{Knapp2024}, discusses how recent developments in heavy lift launch vehicles and satellite mega-constellations in low earth orbit, in conjunction with the upcoming technology demonstrations, make a space-borne radio interferometer more feasible now than ever before.  There are also ongoing efforts to deploy low frequency radio instruments on the lunar surface, described by \cite{Burns2021} and references therein.

The GO-LoW study \cite[]{Knapp2024} identified vector sensors as the best antenna for this application. Vector sensors are advanced antennas consisting of three orthogonal dipoles and three orthogonal loops with a common phase centre. Their benefits include higher sensitivity \cite[]{Kononov2024_ieeeaero} and more degrees of freedom \cite[]{Knapp2016} than the commonly considered dipole triads. However, a space-borne array of vector sensors would be non-coplanar and have a spherical, all-sky field of view, requiring a different imaging process than the traditional Fourier inversion between visibilities and brightness distribution \cite[for example]{thompson_interferometry_2017}.

Ground based dipole arrays have nearly-hemispherical fields-of-view and are nominally co-planar. They can form flattened (i.e., sine projection onto zenith) all-sky images via the traditional Fourier inversion formalism, which is mathematically correct for unpolarized sources \cite[]{sokolowski_2024_high}, but suffers polarization aberration for partially-polarized sources at low elevations \cite[]{carozzi_generalized_2009}. In contrast, the OVRO-LWA \cite[]{eastwood_radio_2018} and the SKA-Low's EDA2 \cite[]{kriele_imaging_2022} form spherical all-sky images using the $m$-mode algorithm proposed by \cite{shaw_coaxing_2015}. Conceptually, the $m$-mode algorithm estimates the spherical harmonic coefficients of the all-sky brightness distribution analogously to how traditional Fourier inversion estimates spatial frequencies of a flat-sky image. Our work applies this concept to a space-borne vector sensor array with non-coplanar motion and a spherical instantaneous field of view.

The rest of this paper is organized in the following way. Section \ref{sec:meas} defines the measurement model for vector sensor interferometry. Section \ref{sec:sph} presents the all-sky imaging method based on spherical harmonics. Section \ref{sec:fisher} describes the use of Fisher information to assess a vector sensor array's harmonic mode sensitivity. Section \ref{sec:sim} demonstrates simulation results. Finally, Section \ref{sec:conc} provides conclusions and future work directions.

%% file: 02_equation.tex
\section{Measurement Model}\label{sec:meas}

\begin{figure}
    \centering
    \includegraphics[alt={Figure with coordinate system on left and polarization ellipse on right.}, width=\linewidth]{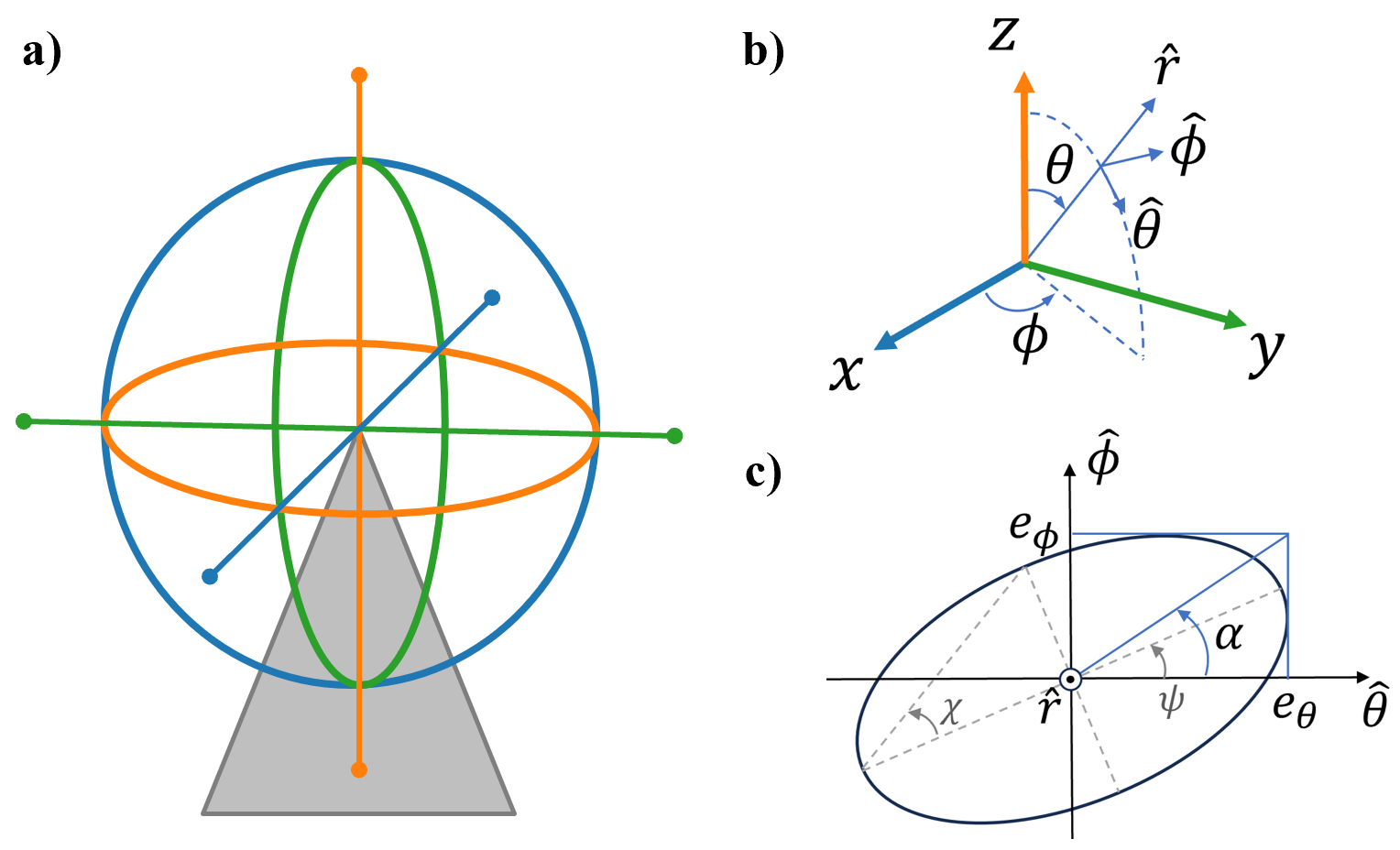}
    \caption{(a) Schematic diagram of a vector sensor, composed of three orthogonal dipole antennas and three orthogonal loop antennas with a common phase centre. The sensor is oriented to align the dipoles with the coordinate axes shown in the next panel, and the loops and dipoles are colour-coded to match the axes with which they align; (b) coordinate system and (c) polarization basis used in this work.}
    \label{fig:coordinate_system}
\end{figure}

The idealized vector sensor, pictured in Figure \ref{fig:coordinate_system}(a), was previously analysed by several authors \cite[]{nehorai_vectorsensor_1994,wong_closedform_2000,fenn_vector_2017}, but all arrived at different transfer functions due to using different coordinate systems, polarization basis, or receive/transmit assumptions. This work considers a receiving vector sensor in the IEEE Std.\ 149 \cite[]{ieee_standard_1979} coordinate system, illustrated in Figure \ref{fig:coordinate_system}(b), and with polarization parametrized by auxiliary angle $\alpha$ and phase difference $\delta$, shown in Figure \ref{fig:coordinate_system}(c).

The polarization response matrix---also known as Jones matrix or voltage beam pattern---of an ideal vector sensor is:
\begin{equation}\label{eq:jones_matrix}
    \bm{A}_\text{vs}(\theta,\phi)=
    \begin{bmatrix}
        -\cos\theta\cos\phi & \sin\phi \\
        -\cos\theta\sin\phi & -\cos\phi \\
        \sin\theta & 0 \\
        -\sin\phi & -\cos\theta\cos\phi \\
        \cos\phi & -\cos\theta\sin\phi \\
        0 & \sin\theta
    \end{bmatrix}
\end{equation}
where the first three rows represent the responses of $x$, $y$, and $z$ dipole elements, and the remaining rows represent the responses of $x$, $y$, and $z$ loop elements. Assuming all elements have unit effective length, the voltage induced onto the vector sensor by an electromagnetic plane wave incident from the direction $(\theta, \phi)$ is:
\begin{equation}\label{eq:vsresponse}
    \bm{v}_\text{vs}(\theta,\phi,\alpha,\delta) = \bm{A}_\text{vs}(\theta,\phi)\bm{e}(\alpha,\delta)
\end{equation}
where $\bm{e}=\bm{\hat{\theta}}\sin{\alpha e^{j\delta}} + \bm{\hat{\phi}}\cos{\alpha}$
is the electric field vector of the incident wave.

The correlation between measurements from these sensors, called a visibility by radio astronomers and a sample covariance matrix in array processing, is found as the time-averaged outer product of voltages at each sensor. For sensors $p$ and $q$, with response matrices $\bm{A}_p$ and $\bm{A}_q$, the visibility is:
\begin{equation}\label{eq:covariance}
    \bm{V}_{pq}=\langle\bm{v}_p\bm{v}_q^H\rangle= \bm{A}_p\langle\bm{ee}^H\rangle\bm{A}_q^H=\bm{A}_p\bm{BA}_q^H,
\end{equation}
where the coherence of the electric field, $\langle\bm{ee}^H\rangle$, is represented by $\bm{B}$, called the brightness matrix. $\bm{B}$ is related to the Stokes parameters via \cite[]{smirnov_revisiting_2011c}:
\begin{equation}
    \bm{B} = \begin{bmatrix}
        I+Q & U+jV \\ U-jV & I-Q
    \end{bmatrix}.
\end{equation}
More generally, a continuous distribution of sources as a function of direction $\hat{\bm{r}}$, incident over the entire $4\pi$ steradians field of view, would produce
\begin{equation}\label{eq:rime_two_element}
    \bm{V}_{pq}=\int_{4\pi}\bm{A}_p(\hat{\bm{r}})\bm{B}(\hat{\bm{r}})\bm{A}_q^H(\hat{\bm{r}})\;\mathrm{d}\Omega,
\end{equation}
where $\mathrm{d}\Omega = \sin\theta\;\mathrm{d}\theta\;\mathrm{d}\phi$ is the solid angle differential. Equation (\ref{eq:rime_two_element}) serves as our measurement model, and next we present an inversion algorithm for obtaining the brightness distribution.

%% file: 03_spherical.tex
\section{Spherical Harmonic Estimation}\label{sec:sph}

Traditional synthesis imaging takes advantage of the Fourier transform relationship between interferometric visibilities and sky brightness distribution, which offers a great computational advantage when inverting measurements to obtain an image. Unfortunately, that method requires flat-sky and coplanarity assumptions, which do not hold for space-borne arrays with all-sky fields of view. Instead, schemes using spherical harmonics as the spatial frequency kernel hold promise to enable computationally-efficient Fourier-like imaging for all-sky cases. Formulation of all-sky interferometric measurements using spherical harmonics has been studied previously in the context of cosmic microwave background experiments \cite[]{kim_direct_2007,mcewen_simulating_2008,liu_spherical_2016,ghosh_deconvolving_2018} and in the context of imaging with ground-based transit radio telescopes \cite[]{shaw_coaxing_2015,eastwood_radio_2018,kriele_imaging_2022}. In this section, we describe a scheme for using spherical harmonics for all-sky imaging with vector sensor interferometers, although this formalism can be applied to any wide field of view antenna hosted on the spacecraft. For further detail on the derivation and implementation, see \cite{kononov_synthesis_2024}.

\subsection{Spherical harmonic formalism}\label{sec:3p1}
We manipulate Equation (\ref{eq:rime_two_element}) to make use of spherical harmonics. The derivation follows the first several steps from the $m$-mode formalism of \cite{shaw_coaxing_2015}. For simplicity, we start with the visibility between a pair of individual feeds---a scalar $V_{pq}$ for feeds $p$ and $q$, which would be one entry in the visibility matrix $\bm{V}_{pq}$ of Equation (\ref{eq:rime_two_element}). First, the polarized brightness distribution $\bm{B}$ is expressed in terms of the Stokes parameters $I$, $Q$, $U$, and $V$:
\begin{equation}
    \bm{B} = \bm{P}_II+\bm{P}_QQ+\bm{P}_UU+\bm{P}_VV,
\end{equation}
where $P_i$ are Pauli matrices:
\begin{align}
    \bm{P}_I&=\begin{bmatrix}1&0\\0&1\end{bmatrix},\quad
    \bm{P}_Q=\begin{bmatrix}1&0\\0&-1\end{bmatrix},\\
    \bm{P}_U&=\begin{bmatrix}0&1\\1&0\end{bmatrix},\quad
    \bm{P}_V=\begin{bmatrix}0&-j\\j&0\end{bmatrix}.
\end{align}
The substitution results in:
\begin{align}
    V_{pq} &= \int_{4\pi}\bm{A}_p(\hat{\bm{r}})\left[\bm{P}_II+\bm{P}_QQ+\bm{P}_UU+\bm{P}_VV\right]\bm{A}_q^H(\hat{\bm{r}})\;\mathrm{d}\Omega \\
    &= \sum_{x\in\{I,Q,U,V\}} \left[\int_{4\pi} \bm{A}_p(\hat{\bm{r}})\bm{P}_x\bm{A}_q^H(\hat{\bm{r}}) T_x(\hat{\bm{r}}) \;\mathrm{d}\Omega\right],\label{eq:eq0}
\end{align}
where $\bm{A}_p$ and $\bm{A}_q$ are now row vectors corresponding to the responses of feeds $p$ and $q$, and
$T_x(\hat{\bm{r}})$ for $x\in\{I,Q,U,V\}$ is a real-valued scalar brightness distribution for each Stokes parameter (\textit{i.e.}, $T_I$ is for Stokes $I$, etc.).
Let $Z_{pqx} = \bm{A}_p\bm{P}_x\bm{A}_q^{H}$ be the complex-valued scalar beam transfer function between feeds $p$ and $q$ for Stokes parameter $x$, and substitute it into (\ref{eq:eq0}):
\begin{equation}\label{eq:eq1}
    V_{pq} = \sum_{x\in\{I,Q,U,V\}} \left[\int_{4\pi} Z_{pqx}(\hat{\bm{r}})T_x(\hat{\bm{r}}) \;\mathrm{d}\Omega\right].
\end{equation}

Next, the polarized brightness temperatures $T_x(\hat{\bm{r}})$ are expanded into spherical harmonics, and the polarized beam transfer functions $Z_{pqx}(\hat{\bm{r}})$ are expanded into conjugate spherical harmonics. The terms corresponding to Stokes parameters $Q$ and $U$, representing linearly polarized components, must be expanded using spin-weighted harmonics because they are not rotationally invariant \cite[]{zaldarriaga_allsky_1997}. For convenience, rotationally invariant quantities $E$ and $B$\footnote{Note that this scalar $B$ is different from the matrix $\bm{B}$ of Equation (\ref{eq:rime_two_element}).} are defined to replace $Q$ and $U$ (see Equations (9)-(16) in \cite{shaw_coaxing_2015} for detail). The harmonic expansions are substituted into Equation (\ref{eq:eq1}). After simplification using the orthogonality property of spherical harmonics, Equation (\ref{eq:eq1}) reduces to:
\begin{equation}\label{eq:eq2}
    V_{pq} = \sum_{x\in\{I,E,B,V\}} \left[\sum_{\ell=0}^\infty \sum_{m=-\ell}^\ell z_{pqx\ell m}a_{x\ell m}\right],
\end{equation}
where $z_{pqx\ell m}$ is the degree $\ell$ order $m$ spherical harmonic coefficient of $Z_{pqx}$, and $a_{x\ell m}$ is the same for $T_x$.

Finally, we consider the visibilities between all possible pairs of feeds in the array and write an equivalent expression to Equation (\ref{eq:eq2}) in matrix notation. The enumerated feed pairs $pq$ become the row index, and the enumerated Stokes/degree/order $x\ell m$ combinations become the column index such that 
\begin{equation}\label{eq:matrix_equation}
    \bm{v} = \bm{Za} + \bm{n}
\end{equation}
where $\bm{v}$ is a column vector of measured visibilities, $\bm{n}$ is a vector of random noise for each visibility, and $\bm{a}$ is a vector of spherical harmonic coefficients representing the sky brightness distribution. The beam transfer matrix $\bm{Z}$ is composed of the spherical harmonic expansion coefficients of the beam transfer functions, which are computed solely based on array geometry. Note that we have to truncate the infinite series harmonic expansions at some $\ell_\text{max}$ in order to construct the matrix expression. The choice of $\ell_\text{max}$ has to be appropriate for the harmonic mode coverage, or spatial band-limit, of the array, which will be discussed in Section \ref{sec:fisher}. An example of the construction of this matrix is given by \cite{kriele_imaging_2022}.

\subsection{Solution with Tikhonov regularization}
The maximum likelihood estimate of the spherical harmonic coefficients, assuming the noise has Gaussian statistics with covariance $\langle\bm{nn}^H\rangle=\bm{N}$, can be found with the standard map-making (linear) solution \cite[]{tegmark_how_1997}:
\begin{equation}\label{eq:ml_solution}
    \bm{a}_{ML} = (\bm{Z}^H\bm{N}^{-1}\bm{Z})^{-1}\bm{Z}^H\bm{N}^{-1}\bm{v}
\end{equation}
The modes which have low Fisher information (\textit{i.e.}, low confidence of estimating them correctly, see Section \ref{sec:fisher}) should be suppressed. One way to do that is Tikhonov regularization \cite[]{eastwood_radio_2018}, also known as ridge regression, which adds a small value $\varepsilon$ to the diagonal of $\bm{Z}^H\bm{N}^{-1}\bm{Z}$ to make the matrix less singular:
\begin{equation}
    \bm{a}_{TR} = (\bm{Z}^H\bm{N}^{-1}\bm{Z}+\varepsilon\bm{I})^{-1}\bm{Z}^H\bm{N}^{-1}\bm{v}
\end{equation}
The value of the ridge parameter $\varepsilon$ is chosen by the practitioner -- $\varepsilon$ too large causes the least-squares residuals to be large, but $\varepsilon$ too small will not adequately suppress the poorly-measured modes and may result in an image that does not represent reality. The optimum $\varepsilon$ lies at the `knee' of an L-curve, such as those presented by \cite{eastwood_radio_2018} or \cite{kriele_imaging_2022}.

\subsection{Point spread function and deconvolution}\label{sec:cleanpsf}
Due to incomplete sampling of the visibility function by the interferometer's baselines, not all spherical harmonic coefficients can be observed. All modes with $\ell > \ell_\text{max}$ are assumed zero, and the modes which have low Fisher information are suppressed with Tikhonov regularization. The point spread function (PSF) that results from the missing and attenuated modes is given by \cite{eastwood_radio_2018} as:
\begin{equation}\label{eq:a_psf}
    \bm{a}_\text{PSF}(\hat{\bm{r}}) = (\bm{Z}^H\bm{Z} + \varepsilon \bm{I})^{-1}{(\bm{Z}^H\bm{Z})}\bm{a}_\text{PS}(\hat{\bm{r}})
\end{equation}
where $\bm{a}_\text{PSF}(\hat{\bm{r}})$ and $\bm{a}_\text{PS}(\hat{\bm{r}})$ are vectors of spherical harmonic coefficients representing the PSF and a point source, respectively, each in the direction $\hat{\bm{r}}$. The spherical harmonic coefficients of a point source are given by
\begin{eqnarray}
    a_{\text{PS},\ell m}(\theta,\phi) = Y_{\ell m}^*(\theta,\phi),
\end{eqnarray}

The computation of $\bm{a}_\text{PSF}$ has to be repeated for each iteration of CLEAN, but there are several ways to accelerate it. One way is to use a look-up table of pre-computed values. When observed from the Earth's surface by a transit telescope, point sources at the same declination take the same track across the sky, and hence have the same PSF. Therefore, deconvolution can be sped up by pre-computing a look-up-table for PSF as a function of declination, and then during deconvolution, the PSF for a given declination is selected and rotated via a simple phase shift to a given longitude \cite[]{eastwood_radio_2018,kriele_imaging_2022}. The longitudinal invariance of the PSF is a feature of transit telescopes only, and is not generalizable to a space-borne array that is not fixed to a rotating platform. If a look-up-table was used for the general case, the PSF would have to be pre-computed over a grid of declinations and longitudes, potentially yielding a prohibitively large table.

Another way to accelerate the computation of $\bm{a}_\text{PSF}$ is discussed by \cite{eastwood_radio_2018}, and is implemented in this work. The matrix $\bm{Z}^H\bm{Z}$ is direction-independent and could be pre-computed. The shape of $\bm{Z}$, introduced at the end of Section \ref{sec:3p1}, is (\textit{number of baselines} $\times$ \textit{number of coefficients}). For arrays such as OVRO-LWA \cite[]{eastwood_radio_2018}, which have more baselines than spherical harmonics being estimated, $\bm{Z}^H\bm{Z}$ yields a smaller matrix than $\bm{Z}$. Vector sensor arrays are likely to fall in the same category, as each pair of sensors creates several unique baselines. Furthermore, the Cholesky decomposition of $\bm{Z}^H\bm{Z} + \varepsilon \bm{I}=\bm{U}^H\bm{U}$ could be pre-computed. $\bm{U}$ is an upper triangular matrix and is inverted more rapidly than a general matrix. Thus, $\bm{a}_\text{PSF}$ can be evaluated as
\begin{equation}
    \bm{a}_\text{PSF}=\bm{U}^{-1}(\bm{U}^H)^{-1}(\bm{Z}^H\bm{Z})\bm{a}_\text{PS}
\end{equation}
Note that since matrix multiplication is associative, we could pre-compute the entire factor in front of $\bm{a}_\text{PS}$. However, that would not be advantageous for large matrices because multiplying a matrix by a matrix takes many more operations than multiplying a vector by a matrix.

%% file: 04_coverage.tex
\section{Harmonic Mode Coverage}\label{sec:fisher}

In traditional (flat-sky) radio interferometers, the spatial frequency coverage is assessed using the $uv$-plane: baselines are plotted in the $uv$-plane, and Earth rotation synthesis fills in the $uv$-plane as Earth's surface rotates with respect to the astronomical target \cite[]{thompson_interferometry_2017}. The locus of points collected in the $uv$-plane shows how well-sampled the visibility space is by the interferometer. In analogue to this, the baselines of all-sky interferometers could be plotted in a $uvw$-volume, but it would be visually intractable and not very useful because the goal in our application is not to image a volume but a spherical surface. We instead assess spherical harmonic mode coverage using Fisher information as an all-sky analogue to the $uv$-plane coverage of flat-sky interferometry.

\subsection{Fisher information}
Consider the measurement model given by Equation (\ref{eq:matrix_equation}), where the parameters $\bm{a}$ and $\bm{n}$ are unknown and treated as vector random variables. The imaging problem is to estimate the parameter $\bm{a}$ given $\bm{v}_0$, a sample of $\bm{v}$. The probability of getting the measurement $\bm{v}_0$ conditioned on $\bm{a}$ is given by a multivariate complex Gaussian density function \cite[\S14.1]{dodelson_modern_2020},
\begin{equation}
    P(\bm{v}_0|\bm{a}) = \frac{1}{\pi^k\det(\bm{N})}e^{-(\bm{v}_0-\bm{Za})^H\bm{N}^{-1}(\bm{v}_0-\bm{Za})},
\end{equation}
where we assume that the array's beam transfer matrix $\bm{Z}$ and the noise covariance matrix $\bm{N}=\langle\bm{nn}^H\rangle$ are known, and $k$ is the number of entries in $\bm{v}_0$. $P(\bm{v}_0|\bm{a})$ is also known as the likelihood function $\mathcal{L}(\bm{a}|\bm{v}_0)$ when viewed as a function of the parameter being estimated rather than a function of the data:
\begin{equation}
    \mathcal{L}(\bm{a}|\bm{v}_0) \equiv P(\bm{v}_0|\bm{a}).
\end{equation}

To build intuition about the likelihood function, suppose that we would like to obtain the probability density of $\bm{a}$ conditioned on the measurement $\bm{v}_0$. Using Bayes rule,
\begin{equation}
    P(\bm{a}|\bm{v}_0) = \frac{P(\bm{v}_0|\bm{a})P(\bm{a})}{P(\bm{v}_0)}.
\end{equation}
The second term in the numerator expresses any prior information we have on the parameter. Assuming we know nothing about $\bm{a}$, we use a uniformly distributed prior, which makes the second term in the numerator a constant. The denominator is independent of the parameter being estimated and serves as a normalization constant. Hence,
\begin{equation}
    P(\bm{a}|\bm{v}_0) \propto P(\bm{v}_0|\bm{a}) = \mathcal{L}(\bm{a}|\bm{v}_0).
\end{equation}
The proportionality constant does not affect the shape of the likelihood function (\textit{i.e.}, the location and width of its peak), so we ignore it for the following analysis.

The log-likelihood function is
\begin{equation}
    \ln \mathcal{L}(\bm{a}|\bm{v}_0) = -\ln\left(\pi^k\det(\bm{N})\right)-(\bm{v}_0-\bm{Za})^H\bm{N}^{-1}(\bm{v}_0-\bm{Za}).
\end{equation}
The Fisher information matrix is defined as the expectation of the log-likelihood's Hessian matrix:
\begin{align}
    \bm{F} &\equiv -\left\langle \frac{\partial^2}{\partial \bm{a} \partial \bm{a}^H} \ln \mathcal{L} \right\rangle \\ &= \bm{Z}^H\bm{N}^{-1}\bm{Z}.
\end{align}
The Fisher information is a measure of the width of the likelihood function's peak \cite[]{tegmark_karhunenloeve_1997}, or in other words, a measure of the confidence in the estimate of $\bm{a}$. We can use $\bm{F}^{-1}$ as an estimate of the covariance $\langle\bm{aa}^H\rangle$, and in particular, the diagonal entries of $\bm{F}^{-1}$ indicate the variance of each element of $\bm{a}$ \cite[]{tegmark_karhunenloeve_1997}. Entries with a low variance indicate harmonic modes that are estimated with high confidence, while entries with a high variance indicate modes estimated with low confidence. \cite{shaw_coaxing_2015} describe using the diagonal elements of $\bm{F}^{-1/2}$ (the standard deviation) as a measure of the amount information a given array can observe about a given harmonic coefficient. We apply this method here to analyse the harmonic mode coverage of vector sensor arrays. 

\begin{figure}
    \centering
    \includegraphics[width=\linewidth]{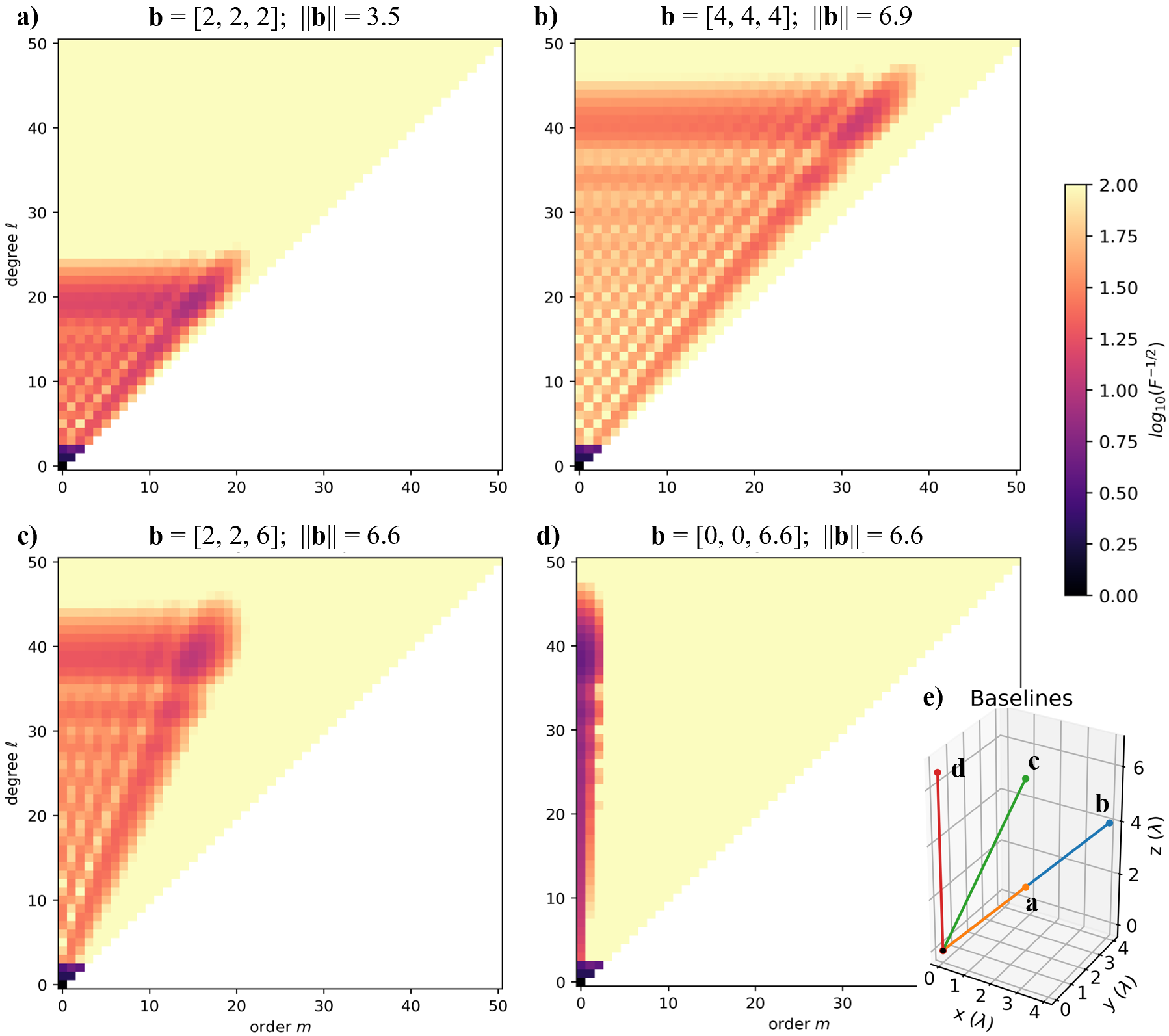}
    \caption[Comparison of harmonic mode coverage between interferometric baseline configurations.] {Comparison of harmonic mode coverage between interferometric baseline configurations for a pair of vector sensors. The baselines used to generate panels (a) and (b) are in same direction but of different lengths, and comparing the two panels shows the effect of changing the baseline length to provide information on higher degree $\ell$ modes. The baselines used to generate panels (b), (c), and (d) all have roughly the same length but different directions, and comparing these panels shows the effect of changing the baseline direction to maximize information provided on specific order $m$ modes within a degree $\ell$ band corresponding to the baseline length. Inset (e) shows the directions and lengths of the baselines presented in (a)--(d).}
    \label{fig:fisher_baselines}
\end{figure}

Figure \ref{fig:fisher_baselines} shows examples of harmonic mode coverages for an interferometric pair of vector sensors with baselines of varying length and direction. All baselines are in units of wavelengths. In Figure \ref{fig:fisher_baselines}, the diagonal entries of $\bm{F}^{-1/2}$, the standard deviation of each harmonic coefficient, are arranged by degree $\ell$ and order $m$, and are plotted in logarithmic scale. The colour scale represents a measure of the uncertainty in a given harmonic mode (\textit{e.g.}, perfect knowledge results in 0 uncertainty). Only $m\geq0$ modes are shown because the sky brightness distribution is a real-valued function whose harmonic coefficients have the property $a_{l(-m)}=(-1)^ma_{\ell m}^*$. Knowledge of an $m>0$ mode gives the exact same knowledge of the corresponding $m<0$ mode. The plots in Figure \ref{fig:fisher_baselines} show that the Fisher information of a beam transfer matrix can be used as a design tool for all-sky interferometers, similarly to how $uv$-plane coverage is used for flat-sky interferometers.

The the highest order harmonic mode that can be measured by an interferometer, $\ell_\text{max}$, is determined by the longest baseline of the interferometer. An estimate of $\ell_\text{max}$ for an array with maximum baseline length $b_\text{max}$ (normalized by wavelength $\lambda$) is given by \cite{shaw_coaxing_2015}:
\begin{equation}\label{eq:lmax}
    \ell_\text{max}\approx2\pi b_\text{max}.
\end{equation}
The plots in Figure \ref{fig:fisher_baselines} show that an individual baseline provides coverage in a band of $\ell$ near $\ell_\text{max}$. For example, the length 3.5$\lambda$ baseline in Figure \ref{fig:fisher_baselines}(a) has $\ell_\text{max}=22$, while the length 6.9$\lambda$ baseline in panel (b) has $\ell_\text{max}=43$. The plots also show that the orientation of the baseline determines the range of $m$ covered. Particularly, $|m|<2\pi\sin\theta_b$, where $\theta_b$ is the elevation angle of the baseline expressed in spherical coordinates \cite[]{shaw_coaxing_2015}. Analogously, there could be an $\ell_\text{min}$ parameter determined by the shortest baseline. However, vector sensor interferometers would always have $\ell_\text{min}=0$ because of the ability to correlate measurements of feeds within a single vector sensor (\textit{i.e.}, a zero-length baseline).

\subsection{Comparing tripoles and vector sensors using Fisher information} \label{sec:comp_vstripole}

\begin{figure}
    \centering
    \includegraphics[width=\linewidth]{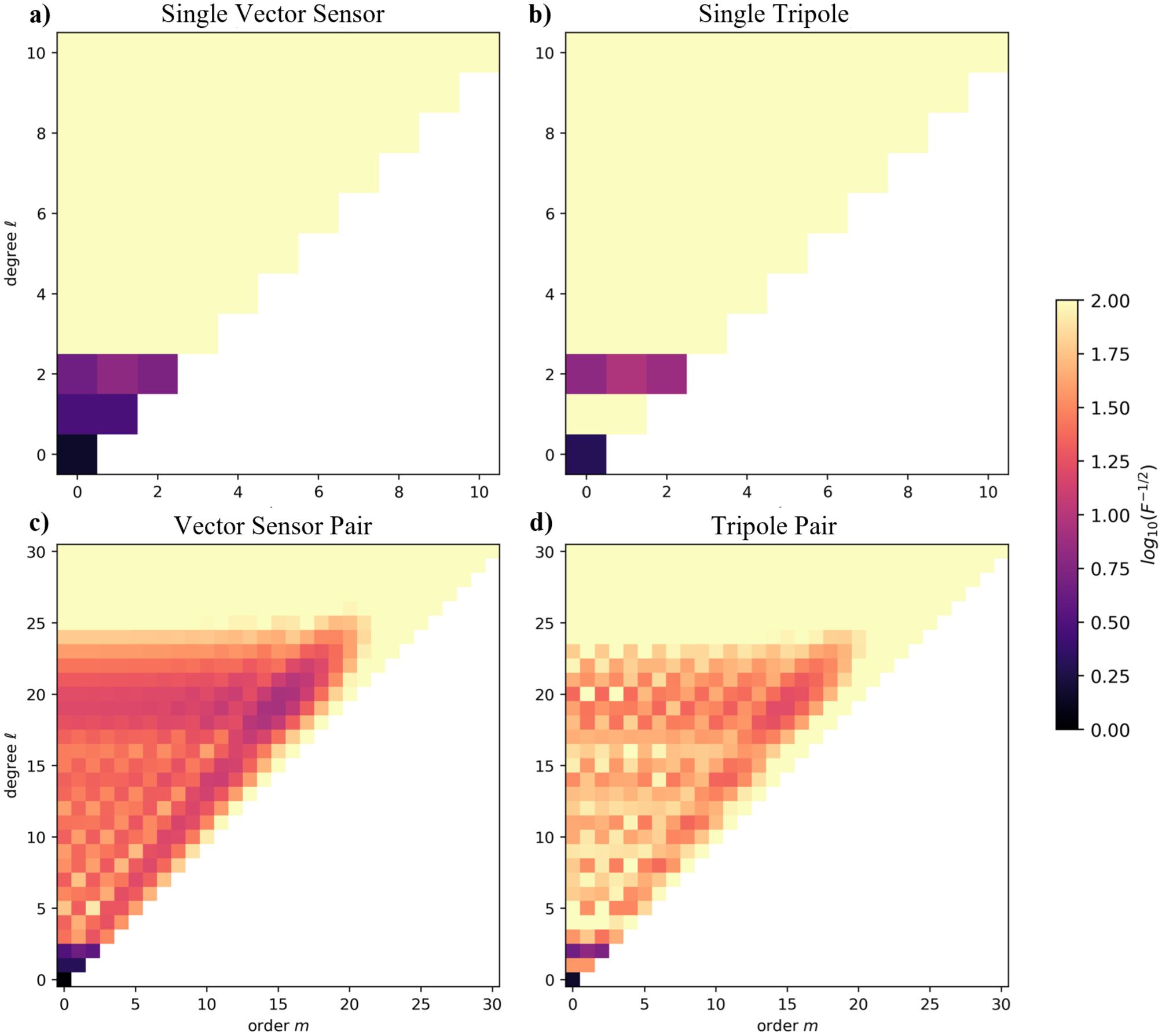}
    \caption[Comparison of harmonic mode coverage between vector sensors and tripoles.] {Comparison of harmonic mode coverage between vector sensors and tripoles. Panels (a) and (b): comparison between a single vector sensor and a single tripole. Panels (c) and (d): comparison between a pair of vector sensors and a pair of tripoles, each separated by a $3.5\lambda$ baseline.}
    \label{fig:fisher_triangles}
\end{figure}

Figure \ref{fig:fisher_triangles} shows a comparison of harmonic mode coverage between vector sensors and tripoles. The upper two panels of Figure \ref{fig:fisher_triangles} show that a single vector sensor can measure the $\ell=0$, $\ell=1$, and $\ell=2$ modes, while a single tripole is blind to the $\ell=1$ modes and has higher uncertainty in the $\ell=2$ modes. The lower two panels of Figure \ref{fig:fisher_triangles} compare interferometric pairs of vector sensors and tripoles, and show that for a pair of sensors, the harmonic mode coverage of a single sensor is still available, but that additional coverage of higher spatial frequency modes is provided by the interferometric baseline. Comparing the two lower panels of Figure \ref{fig:fisher_triangles} to each other makes it clear that the vector sensor pair provides much more information than the tripole pair. This is evidence to support the claim made by \cite{lind_aero_2019} that vector sensors provide additional degrees of freedom to exploit during interferometry, which are not available for sensors with fewer receptors.

This result may seem counter-intuitive because it is well known that the electric and magnetic fields of an electromagnetic plane wave are related, so the vector sensor must be providing redundant information. If only one plane wave is incident onto the sensor, then this is true. However, in an astronomical imaging application, plane waves are incident onto the sensor from all directions simultaneously. The vector sensor measures more distinct linear combinations of these plane waves than a tripole does. That the measurements are distinct is evidenced by the matrix $\bm{A}_\text{vs}$ of Equation (\ref{eq:jones_matrix}). The columns of the first three rows are orthogonal to the respective columns of the last three rows. This is the intuition behind why vector sensors provide more information than tripoles.

\subsection{Total Fisher Information} \label{sec:totalfisher}
Since the spherical harmonic modes are mutually orthogonal, there is no coupling between them, and the diagonal entries of the Fisher information matrix can be summed to produce the total Fisher information available in an array's measurements \cite[]{shaw_coaxing_2015}:
    \begin{equation}
        F_\text{total} = \text{tr}(\bm{F})
    \end{equation}
We use the total Fisher information metric to observe the amount of information measured by an array as the number of sensors in the array grows, and compare the growth rates between arrays of vector sensors and arrays of tripoles.

As we showed in Figure \ref{fig:fisher_baselines}, the information collected by an array depends on the geometric layout of the array. In order to make the comparison independent of array geometry, a Monte Carlo method was used to generate random array geometries. In each trial, a given number of sensors was placed in random positions drawn uniformly from a sphere of radius 10 wavelengths, and the total Fisher information through $\ell_\text{max}=180$ was calculated. The parameter $\ell_\text{max}$ was chosen slightly larger than required by the longest baseline to ensure that all the information was included in the total. 200 trials were performed for each type of sensor (vector and tripole) and each number of sensors.

\begin{figure}
    \centering
    \includegraphics[width=0.9\linewidth]{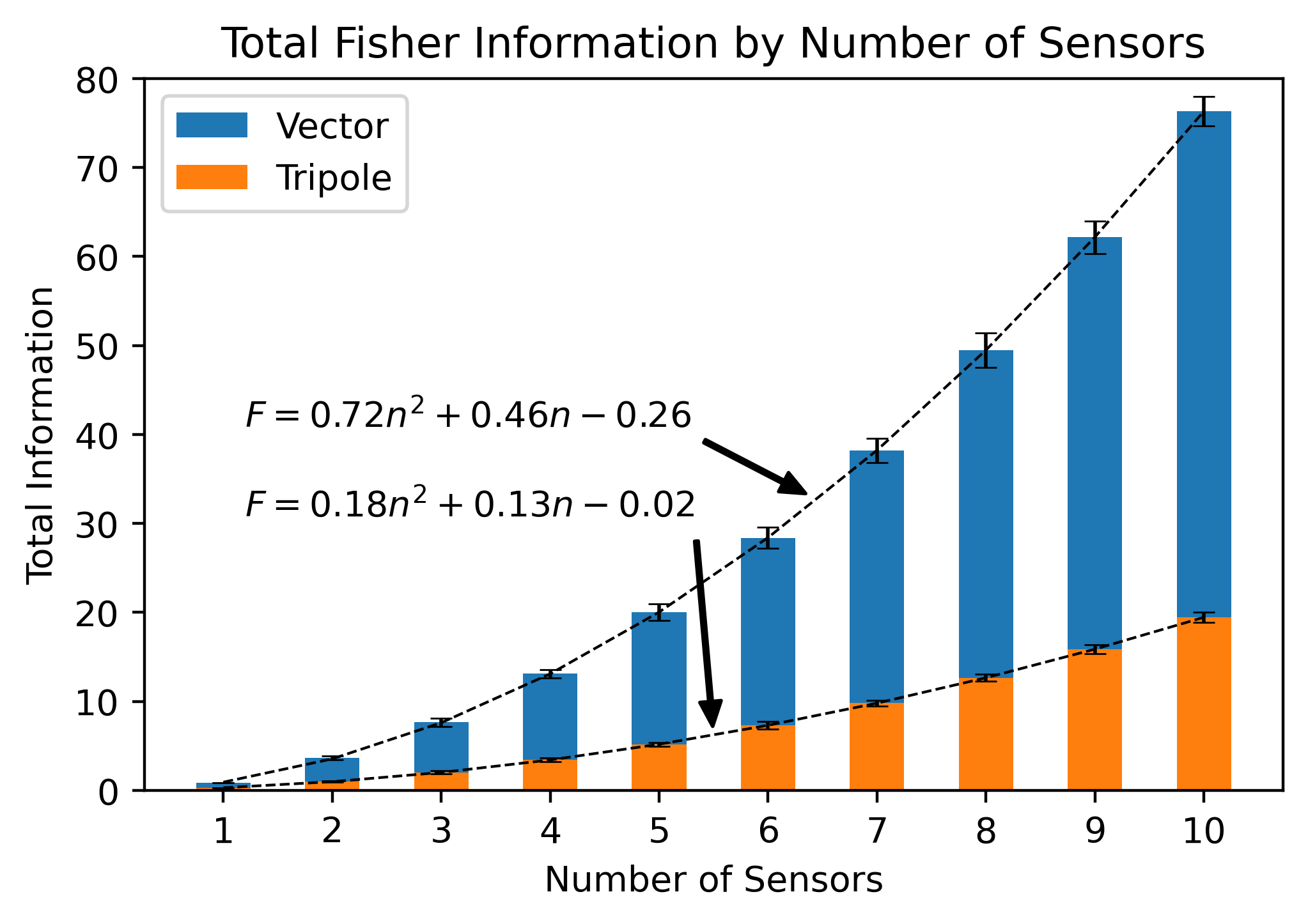}
    \caption[Comparison of total Fisher information versus number of nodes between vector sensor and tripole arrays.] {Comparison of total Fisher information versus number of nodes between vector sensor and tripole arrays. The heights of the bars indicate the mean from 200 Monte Carlo trials, while error bars indicate 10 standard deviations. Vector sensor arrays provide 4 times more information than tripole arrays for a given number of nodes.}
    \label{fig:bargraph}
\end{figure}

Figure \ref{fig:bargraph} plots the total Fisher information versus number of sensors for vector sensor and tripole arrays. Error bars indicate 10 standard deviations---instead of one standard deviation---to be visible on the plot. The small standard deviations suggest that the total Fisher information measured by a randomly-spaced array with a given number of sensors is independent of the array geometry. While different array geometries measure different sets of harmonic modes, the total information measured is the same for a given number of sensors in the array, barring flat, co-linear, or regular geometries with many redundant baselines. A quadratic function fits the results for both sensor types very well, and is justified because the number of baselines in an interferometer is proportional to the square of the number of sensors. The ratio of the quadratic trends indicates that vector sensor arrays provide 4 times more information than tripole arrays for a fixed number of nodes. This can be explained because each baseline between a pair of vector sensors provides $36=6^2$ visibilities, which is 4 times greater than the $9=3^2$ visibilities provided by baselines between pairs of tripoles. Finally, it seems like there should be a saturation point where additional nodes provide diminishing returns in terms of new information and the quadratic growth trend ceases. However, this conjecture was not verified experimentally due to computational limitations and is outside of the scope of this work.

%% file: 05_sim.tex
\section{Simulation Results}\label{sec:sim}
To demonstrate the techniques described above, we present an imaging simulation for all-sky mapping.  This imaging simulation uses nodes with vector sensor antennas (6 feeds: three dipoles, three loop antennas) and assumes that the nodes are in space with no obscuration from large bodies, far from the Earth, Moon, and Sun.

\subsection{Input sky map} \label{sec:skymap}
We selected the de-striped Haslam 408~MHz map \cite[]{Haslam1982,remazeilles_improved_2015} as the basis for simulated imaging because it is complete for the whole sky and includes sources at a range of spatial scales, from point sources to the galactic disk and the North Polar Spur.  The 2014 reprocessed map in HEALPix format is available from NASA\footnote{\url{https://lambda.gsfc.nasa.gov/product/foreground/fg_2014_haslam_408_get.html}}. For this simulation, we use a nominal frequency of 1~MHz for array and baseline calculations, although we note that the real 1~MHz sky will look different from the 408~MHz Haslam map. We assume a quasi-monochromatic sky (no bandwidth synthesis) for the purposes of this demonstration.

\begin{figure}
    \centering
    \includegraphics[width=0.9\linewidth]{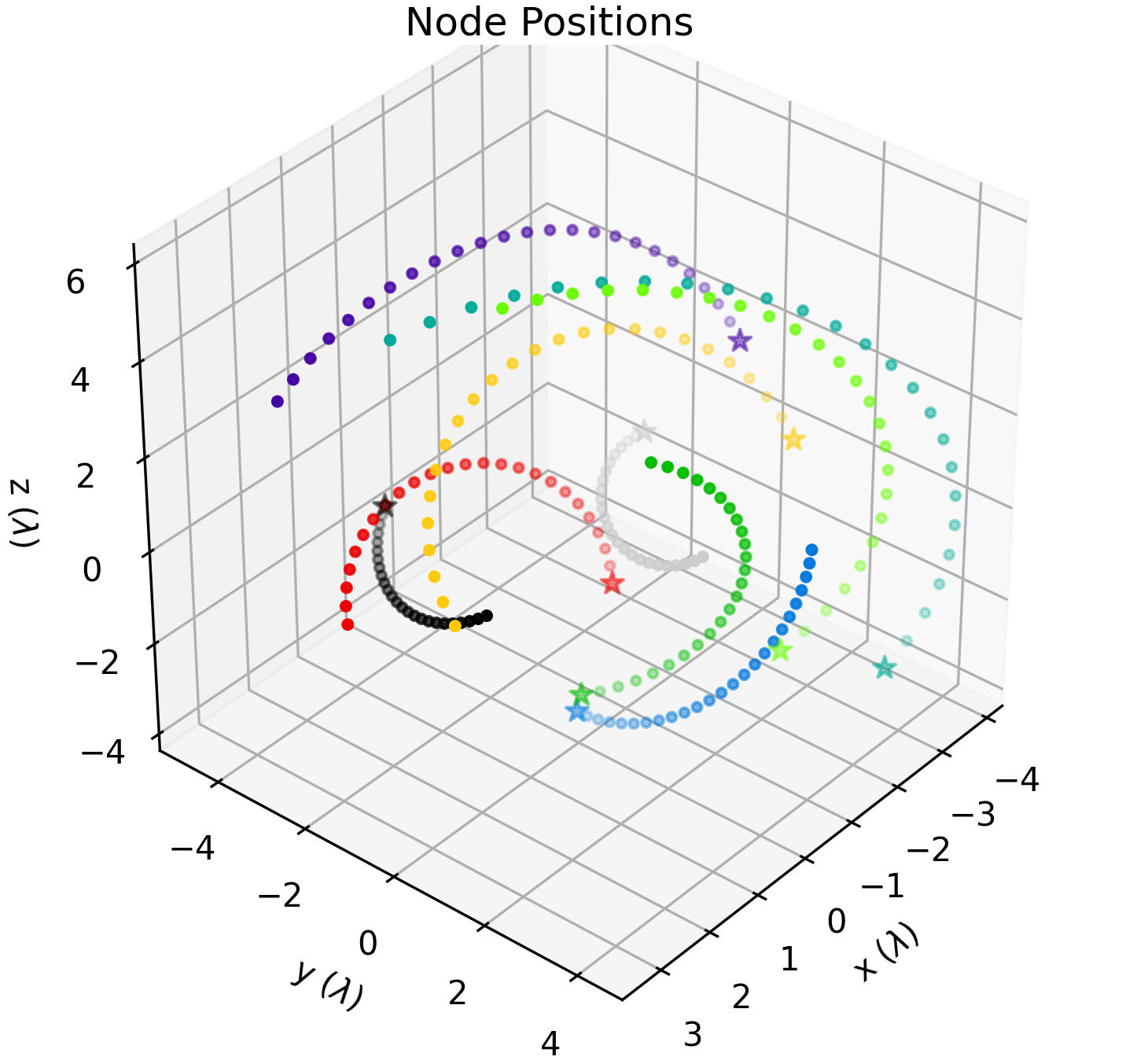}
    \caption{Node locations over rotation synthesis period.  The dotted trail, with a different color representing each unique node, shows node position evolution over time with star markers indicating the initial position of each node. The markers darken with each timestep, indicating the trajectory of each node. The initial positions of each spacecraft are used to form the snapshot baselines in Figure \ref{fig:big_figure}(a).}
    \label{fig:node_rotsynth}
\end{figure}

\begin{figure*}
    \centering
    \includegraphics[width=\textwidth]{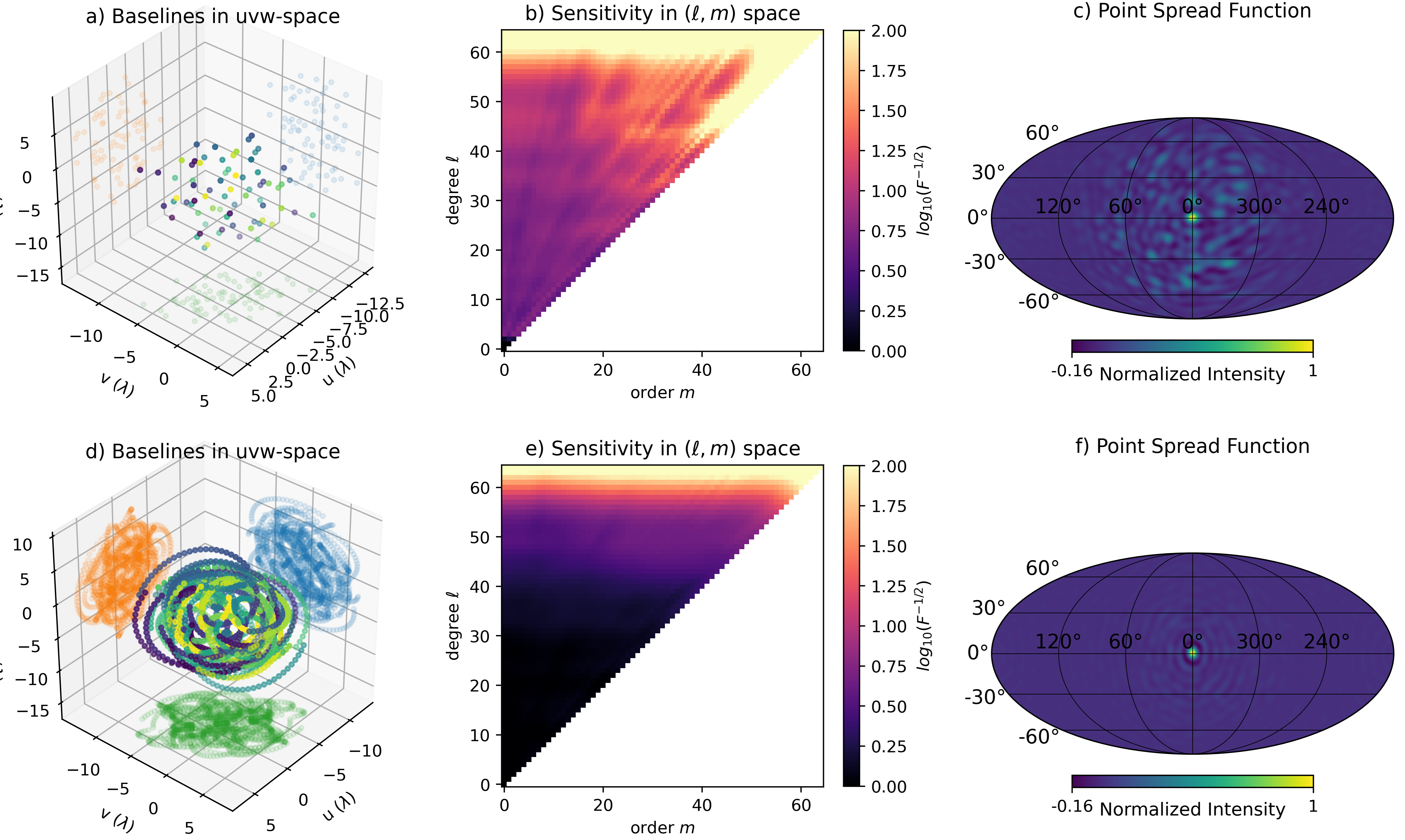}
    \caption[Simulated array characteristics] {Simulated array characteristics for snapshot (top row) and rotation synthesis (bottom row) imaging. Panels (a) and (d): Baselines corresponding to the node positions of Figure \ref{fig:node_rotsynth}.  Projections in principle axis planes are shown for reference. Panels (b) and (e): Harmonic mode coverage of the array.  The harmonic mode coverage represented in (e) includes multiple baselines for each antenna pair that are generated as the nodes move. Compared with (b), rotation synthesis increases the certainty in estimates of many more spherical harmonic components---up to approximately $\ell_\text{max}=40$---as intended with this array design (see Section \ref{sect:simarray}). Panels (c) and (f): Point spread functions.}
    \label{fig:big_figure}
\end{figure*}

\subsection{Simulated array}\label{sect:simarray}
The simulated interferometric array is composed of nine receiving spacecraft, each with a vector sensor antenna.  The spacecraft, referred to as nodes, are assumed to be in free space with 4$\pi$ steradian fields of view.  

The spacing of the nodes is set by choosing a maximum separation that would be appropriate for estimating spherical harmonic coefficients up to $\ell_\text{max}=40$ based on the relationship in Equation (\ref{eq:lmax}) with $\lambda = 300$~m (1~MHz).  With a 1.9~km spacing as an upper bound and 0.5~km as the lower bound, we chose nine initial positions at random by drawing from a uniform distribution.  We chose a goal of imaging to $\ell_\text{max}=40$ as a compromise between image fidelity and computational burden.

The nodes are moved along 3D trajectories for 25 time-steps in order to obtain multiple snapshots with different baseline configurations. We refer to this as rotation synthesis, analogous to Earth rotation synthesis but without the constraint that the receivers are fixed to a surface. The motion described is a spatial translation only; the antenna elements maintain the same orientation relative to the coordinate frame throughout the simulation.  The goal is to simulate a cluster of spacecraft moving on a variety of trajectories in free space while maintaining a three-axis stabilized attitude, as they might to maintain solar panel pointing at the sun. The configuration and motion of the nodes is not based on motion in any particular gravitational environment or orbit; it is intended only to demonstrate the effect of rotation synthesis on the spherical harmonic imaging described in Sections \ref{sec:sph} and \ref{sec:fisher}. Modelling actual orbital dynamics as part of an imaging simulation is the subject of future work.

The baselines and harmonic mode coverage for a single snapshot with nine nodes are shown in Figure \ref{fig:big_figure}(a, b) and for the multi-snapshot rotational synthesis array in Figure \ref{fig:big_figure}(d, e). The harmonic mode coverage resulting from the array with rotational synthesis shows improved ability to estimate spherical harmonic coefficients up to degree $\ell=40$. The sensitivity to harmonics of degree $\ell>40$ worsens rapidly due to the absence of baselines longer than our simulation's upper bound. And lastly, the array's PSF for a single snapshot and with rotation synthesis are calculated as described in Section \ref{sec:cleanpsf}, and pictured in Figures \ref{fig:big_figure}(c) and \ref{fig:big_figure}(f), respectively. While the two PSFs have similar main beam widths, rotational synthesis reduces the sidelobes.

\subsection{Spherical harmonic imaging}
\begin{figure*}
    \centering
    \includegraphics[width=0.95\textwidth]{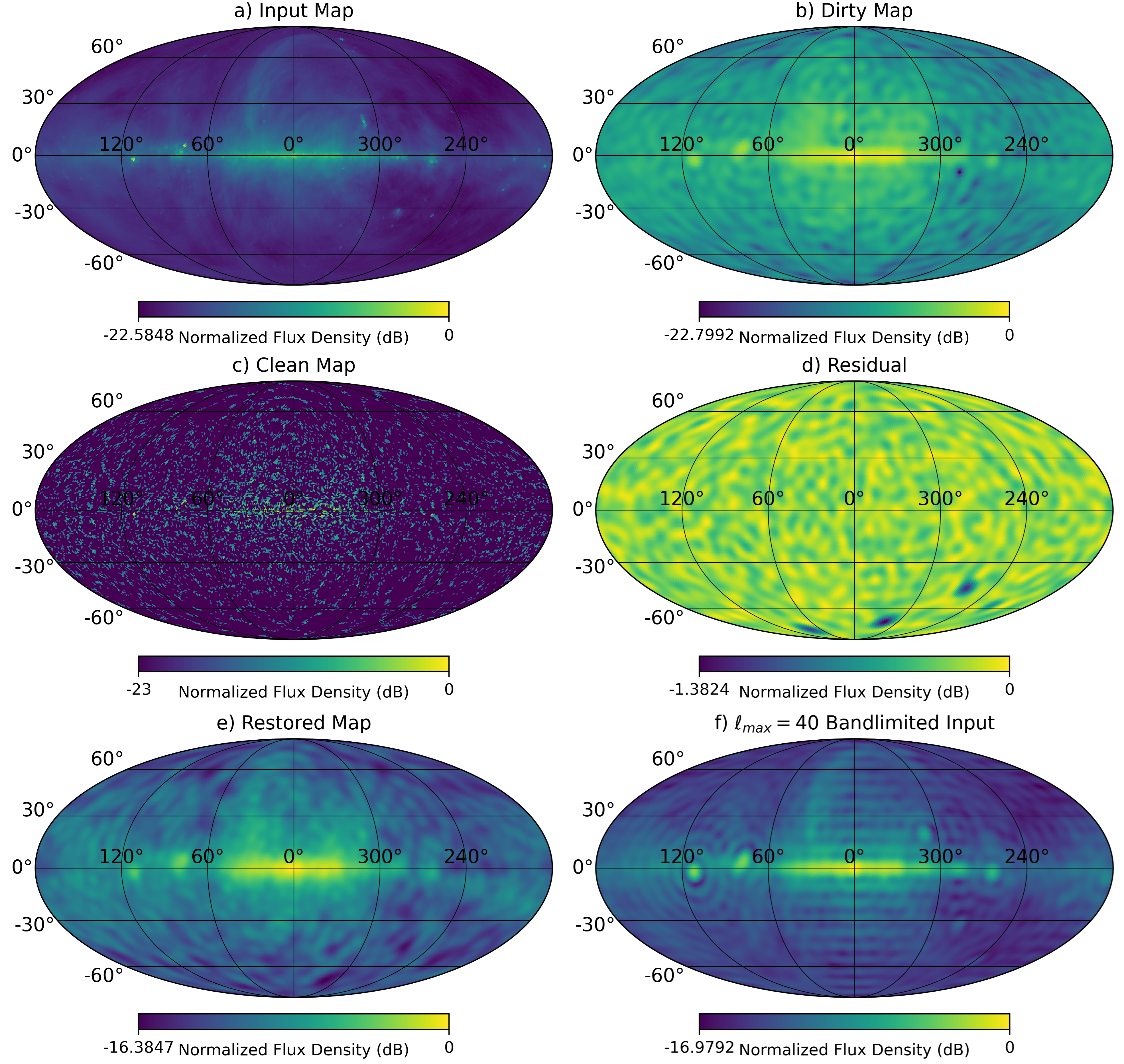}
    \caption[Spherical imaging results] {Spherical imaging results.  Panel (a) shows the input map in logarithmic scale.  Panel (b) is the result of spherical harmonic imaging prior to cleaning (the dirty map).  Panel (c) shows the clean components recovered during cleaning while panel (d) shows the residual map after the PSF (Figure \ref{fig:big_figure}(f)) corresponding to each clean component is subtracted.  Panel (e) shows the restored map after cleaning and the use of a Gaussian restoring beam of 7 degrees.  For comparison, (f) shows the theoretical prediction, the input map of (a) spatially band limited to $\ell_\text{max}=40$, which is the range within which the array is designed to produce high quality spherical harmonic component estimates (see Figure \ref{fig:big_figure}(e)).}
    \label{fig:result_six_panel}
\end{figure*}

The spherical harmonic imaging algorithm (Section \ref{sec:sph}) was applied to the simulated measurements. Figure \ref{fig:result_six_panel} displays results for each stage of the imaging process. Panel (a) shows the input image. Panel (b) shows the dirty map, which contains sources from the input image convolved with the point spread function of the array, and also distorted by errors in estimates of the harmonic coefficients. We performed 10000 iterations of the spherical CLEAN algorithm described by \cite{eastwood_radio_2018} and \cite{kriele_imaging_2022} using Tikhonov regularization $\varepsilon=1$, and loop gain $\gamma=0.1$. The clean map and residual are shown in panels (c) and (d) of Figure \ref{fig:result_six_panel}. Finally, we applied a Gaussian restoring beam of 7$^\circ$, same as the full-width half-maximum of the PSF, and added the residual back, to produce the restored map shown in panel (e).

We measure success by comparing the result of the simulation, Figure \ref{fig:result_six_panel}(e), to the theoretical prediction of the image with the $\ell$ truncated to $\ell_\text{max}=40$, shown in Figure \ref{fig:result_six_panel}(f). We truncate the theoretical prediction at $\ell_\text{max}=40$ because the simulation estimated the first 40 orders of spherical harmonic coefficients (861 total coefficients). If the estimate was exactly correct, then \ref{fig:result_six_panel}(f) shows the result we would obtain. However, while the result is fairly close to the ideal, it is degraded by the sparsity of the aperture and remaining noise in the residual.  Remaining noise is the result of imperfect deconvolution, and measurement noise is not considered in the simulation.

%% file: 06_conclusion.tex
\section{Conclusion}\label{sec:conc}
We defined a measurement model for an electrically small vector sensor (Section \ref{sec:meas}) and a spherical harmonic mode estimation algorithm for imaging with an array of such sensors (Section \ref{sec:sph}). The algorithm can also be employed for other electrically small antennas with appropriate modifications to the measurement model. Our method is tailored to a space-borne scenario where the receiving nodes have an unobstructed $4\pi$ steradian field of view.  This approach builds on existing literature for ground-based interferometric telescopes, but is novel in its application to space-based interferometric arrays.  

Fisher information is used to explore the capabilities of three-dimensional interferometric arrays and compare their performance based on the type of antenna used, tripole or vector sensor (Section \ref{sec:fisher}).  We showed that a single vector sensor can estimate more spherical harmonic components than a tripole antenna, confirming the prior result that vector sensors offer advantages over tripoles in free space \cite[]{Knapp2016, Kononov2024_ieeeaero}.  We also quantified the difference between arrays composed of tripoles and vector sensors in terms of estimated spherical harmonic components, and showed that the information content grows four times faster with number of elements for vector sensor arrays compared to tripole arrays (Figure \ref{fig:bargraph}).

Finally, we demonstrated the spherical harmonic imaging algorithm using a real sky map as input (Section \ref{sec:sim}).  This simulation also showed the effect of rotation synthesis (Figure \ref{fig:big_figure}) in three ways: by plotting baselines in the $uvw$ volume, comparing harmonic mode coverage, and comparing snapshot and synthesis PSFs. The results of the imaging simulation, including the deconvolution of the PSF from the sky brightness, are presented in Figure \ref{fig:result_six_panel}.

To further explore space interferometer all-sky imaging, we are developing the Great Observatory for Long Wavelengths (GO-LoW) concept \cite[]{Knapp2024}, a mega-constellation low frequency interferometric array, under NASA NIAC Phase II funding.  Ongoing and future work on the GO-LoW study features an integrated simulation of orbital dynamics, data reduction, and imaging, with plans to scale the simulation to thousands of nodes. The work presented here will be further refined and expanded to support that study.